\documentstyle [12pt,epsf]{article}
\newcommand{\be}{\begin{equation}}
\newcommand{\ee}{\end{equation}}
\newcommand{\bea}{\begin{eqnarray}}
\newcommand{\eea}{\end{eqnarray}}
\newcommand{\de}{\partial}
\newcommand{\nn}{\nonumber}

\newcommand{\inner}[2]{\big{<}#1\big{|}#2\big{>}}
\newcommand{\da}{\dagger}

\newcommand{\refpa}[1]{(\ref{#1})}
\newcommand{\intll}[3]{\int _#1^#2 d\! #3 \,} 
\newcommand{\gh}{g \hbar}

\font\zz=cmss10
\newcommand{\Z}{\hbox{\zz Z} \kern-.4em \hbox{\zz Z}}
\begin{document}
\begin{titlepage}
\begin{flushleft}
G\"oteborg\\
ITP 93-39\\
gr-qc/9312013\\
November 1993\\
\end{flushleft}
\vspace{1cm}
\begin{center}
\leavevmode
\epsfbox{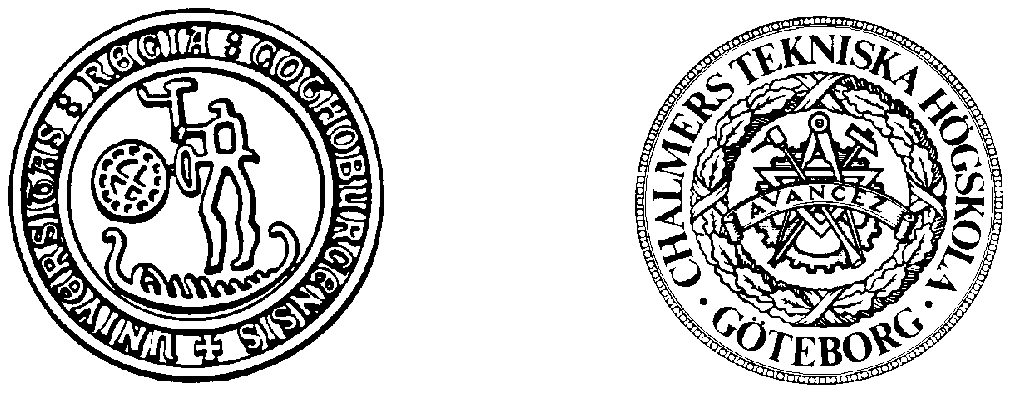}
\end{center}
\vspace{1cm}
\begin{center}
{\Large Representations of the $SU(N)$ $T$-algebra and the loop representation
in $1+1$-dimensions}\\
\vspace{5mm}
{\large Joakim Hallin}\footnote{Email address: tfejh@fy.chalmers.se}\\
\vspace{1cm}
{\sl Institute of Theoretical Physics\\
Chalmers University of Technology\\
and University of G\"oteborg\\
S-412 96 G\"oteborg, Sweden}\\
\vspace{1.5cm}
{\bf Abstract}\\
\end{center}
We consider the phase-space of Yang-Mills on a cylindrical space-time ($S^1
\times {\bf R}$) and the associated algebra of
gauge-invariant
functions, the $T$-variables. We solve the Mandelstam identities both
classically and quantum-mechanically by considering
the $T$-variables as functions of the eigenvalues of the holonomy and their
associated momenta. It is shown
that there are two inequivalent representations of the quantum $T$-algebra.
Then we compare this reduced phase space approach
to Dirac quantization and find it to give essentially equivalent results. We
proceed to define a loop representation
in each of these two cases. One of these loop representations (for $N=2$) is
more or less equivalent to the usual
loop representation.
\end{titlepage}

\section{Introduction}
In \cite{aa1,aa2} Ashtekar showed that the phase-space of pure gravity can be
embedded in the phase-space of
$SL(2,{\bf C})$ Yang-Mills theory. This was used in \cite{rs:qgr} where a
number of gauge-invariant observables for
Yang-Mills theory, the $T$-variables, were used in an attempt to quantize
gravity non-perturbatively. The $T^0$-variable
is the trace of the holonomy of the connection around a loop i.e. what is more
commonly known as the Wilson loop
variable. The higher $T$-variables are phase-space generalizations of $T^0$
also containing momenta, being essentially
time derivatives of $T^0$. This quantization was performed by means of the so
called loop representation where states
are functionals of loops and the $T$-variables act in a specified way on such
states. States satisfying all constraints
were found in this formalism.
However, in \cite{marolf} quantum general relativity in $2+1$-dimensions, which
is embedded in the phase space of
$SL(2,{\bf R})$ Yang-Mills theory, was investigated
and it was found that the loop representation inadequately described the
theory. The reason for this was essentially the
non-compactness of the gauge group. Also in \cite{loll} it was argued that the
loop representation is incomplete due to
the appearance of certain non-linear inequalities being satisfied by the
classical $T$-variables which seemingly
where ignored by the loop representation. A similar scenario is expected to
take place in $3+1$-dimensions although
much less is known there, partly because of the complicated reality conditions
one needs to impose on the phase space
variables.
One possible conclusion one might draw from this is to forget about
$T$-variables. This is however not a very useful
conclusion since after all the $T$-variables form a large class of
gauge-invariant functions on phase space. In this
paper we will, as a toy model for the higher dimensional cases, investigate the
phase space of $SU(N)$ Yang-Mills
theory in $1+1$-dimensions and its corresponding $T$-variables. The motivation
for doing
the  analysis in $1+1$-dimensions is that everything is completely and
explicitly doable. Also, we investigate $SU(N)$
instead of just $SU(2)$ to see whether $SU(2)$ is special in any way. It turns
out that it is not. We will however not
choose the loop
representation as a starting point. Instead we quantize the $T$-algebra by
means of a reduced phase space approach
as well as by Dirac quantization. Then by using the eigenstates of the
Yang-Mills Hamiltonian we can proceed to
define loop representations. These loop representations are
more or less equivalent to the usual loop representation. This is to be
expected since $SU(N)$ is a compact group. In a
forthcoming paper we hope to do the same analysis for $SL(2,{\bf R})$ as we do
here for $SU(N)$.

\section{Classical theory}
Our starting point is the gauge-invariant part of the $SU(N)$ Yang-Mills
Hamiltonian,
\be
H=\frac{1}{2}\intll{0}{L}{x} \mbox{tr}(E^2(x)),
\ee
where $L$ is the length of the circle.
The basic Poisson bracket is,
\be
\{ A^a(x),E^b(y) \}=\delta ^{ab} \delta (x-y),\label{poisson}
\ee
and $A(x)=A^a(x) t^a$, $E(x)=E^a(x) t^a$ where $t^a$ are the $N\times N$-matrix
generators of the group ($a=1,\ldots ,
\mbox{dim}(SU(N))$). Here we have chosen
\[ \mbox{tr}(t^a t^b)=\delta ^{ab} \]
which implies the identity,
\be
(t^a)_{ij} (t^a)_{kl}=\delta_{il} \delta_{jk}-\frac{1}{N} \delta _{ij} \delta
_{kl},\label{eq:fierz}
\ee
where $i,j,k,l$ denotes matrix indices. We assume that the connection and
electric field are periodic fields on the
circle i.e. $A(x)=A(x+L)$, $E(x)=E(x+L)$.
We also have the first class constraint (Gauss' law),
\be
D_x E(x)=\de _x E(x)+ig \lbrack A(x),E(x) \rbrack \approx 0 ,\label{eq:gauss}
\ee
where $g$ is the coupling constant. Let us define parallel transport by
\be
U(x,y)={\cal{P}}\exp (ig \intll{x}{y}{x'}A(x')),
\ee
where $\cal{P}$ denotes path ordering, i.e. $U(x,y)$ is the solution to the
integral equation
\[
U(x,y)=1+ig \intll{x}{y}{x'}U(x,x') A(x').
\]
$U(x,y)$ is an element of the group $SU(N)$.
The holonomy $h(x)$ is defined by
\[ h(x)=U(x,x+L) \]
i.e. parallel transport once around the whole circle. Note also that $U(x,x+n
L)=h^n(x)$ where $n$ is an integer which
follows from the basic sewing property of path ordered exponentials,
\[ U(x,x') U(x',y)=U(x,y).\]
Let $\Lambda (x)$ be a (finite) $SU(N)$ gauge-transformation (generated by
\refpa{eq:gauss}). Then
\bea
A'(x) &=& \Lambda (x) A(x) \Lambda ^{-1}(x)+\frac{1}{ig}\Lambda (x) \de _x
\Lambda ^{-1}(x) \label{gtr}\\
E'(x) &=& \Lambda (x) E(x) \Lambda^{-1}(x).
\eea
This implies that $U(x,y)$ transforms homogeneously i.e.
\be
U'(x,y)=\Lambda (x) U(x,y) \Lambda ^{-1}(y),
\ee
and in particular,
\be
h'(x)=\Lambda (x) h(x) \Lambda ^{-1}(x),\label{eq:hol}
\ee
since $\Lambda (x)$ is periodic. We might allow for non-periodic
gauge-transformations (these are not generated by
the constraint) still keeping $A$ periodic.
These satisfy,
\be
\Lambda (x+L)=\Z _N \, \Lambda (x),
\ee
where $\Z _N$ is any element in the center of the group i.e. an $N$:th root of
unity $\xi$, $\xi ^N=1$. We will call such
non-periodic gauge-transformations $\Z _N$ transformations. Under such a
transformation, the holonomy transforms as,
\be
h'(x)=\xi \, \Lambda (x) h(x) \Lambda ^{-1}(x).
\ee
There is no reason to demand invariance under such transformations unless one
is really interested in one
of the corresponding groups like e.g. $SU(2)/\Z _2 \approx SO(3)$.
Furthermore, as soon as we couple fermions these transformations are not
allowed any longer.
\subsection{Loop variables}
Following \cite{rs:qgr} we introduce the following functions on phase space,
the loop variables,
\bea
T^0(n) &=& \mbox{tr}(h^n(x))\\
T^1(x;n) &=& \mbox{tr}(E(x) h^n(x))\\
T^2(x;n,y;m) &=& \mbox{tr}(E(x)U(x,y+nL)E(y)U(y,x+mL)).
\eea
They are easily seen to be gauge-invariant. Furthermore, $T^0(nN)$ etc, is $\Z
_N$ invariant.
Note also that $T^0(n)$ is independent of $x$, motivating the notation. In
fact, on the constraint surface $T^1(x;n)$ is also independent of $x$ since
\be
\de _x T^1(x;n)=\mbox{tr}((D_x E(x)) h^n(x))\approx 0 .\label{eq:t1}
\ee
Similarly, $T^2(x;n,y;m)$ is independent of $x$ and $y$. Using the identity,
\[ T^2(x+n' L;n,y;m)=T^2(x;n-n',y;m+n'), \]
it also follows that $T^2(x;n,y;m)$ is independent of $n-m$ on the constraint
surface, i.e. $T^2(x;n,y;m)=T^2(n+m)$.
Analogously, one may consider loop variables of higher order
in $E$ i.e.
\[ T^p(x_1;n_1,\ldots ,x_p;n_p).\]
On the constraint surface $T^p$ will only depend on $n_1+\cdots +n_p$. To
calculate Poisson brackets we need,
\be
\frac{\delta U(x,y)}{\delta A^a(x')}=ig\theta (x,y,x')U(x,x')t^aU(x',y),
\ee
where
\[ \theta (x,y,x')=\intll{x}{y}{x''}\delta(x''-x').\]
In particular,
\be
\frac{\delta h(x)}{\delta A^a(x')}=igU(x,x')t^aU(x',x)h(x).\label{eq:func}
\ee
In what follows, all brackets will be evaluated on the constraint surface,
where they simplify. Using \refpa{eq:func}
and \refpa{eq:fierz} we obtain,
\bea
\{ T^0(n),T^0(m) \} &=& 0\nn \\
\{ T^1(n),T^0(m) \} &=& -igm (T^0(n+m)-\frac{1}{N}T^0(n) T^0(m))\label{t1t0} \\
\{ T^1(n),T^1(m) \} &=& ig(n-m) T^1(n+m)-\frac{ig}{N}(n T^1(n) T^0(m)-m
T^1(m)T^0(n))\label{t1t1} \\
\{ T^2(n),T^0(m) \} &=& -2igm (T^1(n+m)-\frac{1}{N}T^1(n) T^0(m))\label{t2t0}\\
\{ T^2(n),T^1(m) \} &=& ig(n-2m)
T^2(n+m)+\frac{ig}{N}(2mT^1(n)T^1(m)-nT^2(n)T^0(m))\label{t2t1} \\
\{ H,T^p(n) \} &=& -ignL\, T^{p+1}(n)\label{htp}.
\eea
The last identity follows since $H=\frac{L}{2}T^2(0)$. Note the central
importance of $H$ (or $T^2(0)$). Even if we're not
interested in the Yang-Mills time-evolution, $H$ still acts as a generator of
$T^p$:s through \refpa{htp}. Since $T^p(n)$ is
a continuos function of $n$, \refpa{htp} defines $T^{p+1}(n)$ even for $n=0$.
We also have the reality conditions,
\bea
(T^p(n))^* &=& T^p(-n) \nn \\
H^*=H,\label{eq:reality}
\eea
where $*$ denotes complex conjugation.

\subsection{Mandelstam identities}
The Mandelstam identities were first discussed for a special case in
\cite{m:cm}. In \cite{rg} and \cite{gt} they are
discussed in general. We will not need to know their general form, we will only
illustrate them by examples. For $N=2$ the
identities are,
\bea
T^0(n) &=& T^0(-n),\label{eq:t01}\\
T^0(n) T^0(m) &=& T^0(n+m)+T^0(n-m)\label{eq:t02},
\eea
while for $N=3$ one of the identities is,
\be
T^0(n)^3-3T^0(n) T^0(2n)+2T^0(3n)=6.
\ee
The other identity for $N=3$ is much more complicated. It is only for $N=2$
that these identities have a simple form. There are
also analogous identities for the higher $T$-variables. Let us illustrate this
for $N=2$. Calculating the bracket of both
the left- and right-hand side of \refpa{eq:t01} with $H$ using \refpa{htp} one
finds,
\be
T^1(n)=-T^1(-n)\label{t1n}.
\ee
Doing the same operation on \refpa{eq:t02} and using \refpa{t1n} one obtains,
\be
n f(n,m)+m f(m,n)=0\label{eq:f1f1},
\ee
where
\be
f(n,m)=T^1(n) T^0(m)-T^1(n+m)-T^1(n-m).
\ee
Using \refpa{eq:t01} and \refpa{t1n} we get $f(n,m)=-f(-n,-m)$. Having found
this it is easy to see that \refpa{eq:f1f1}
implies
$f(n,m)=0$ i.e.,
\be
T^1(n) T^0(m)=T^1(n+m)+T^1(n-m).
\ee
One can then repeat this construction for the $T^1$-identities and construct
identities involving $T^2$ etc. Analogously,
having the explicit form of the identities for $T^0$ for any $N$, one can find
identities involving the higher $T$-variables.

\subsection{Conjugacy classes}
As seen by \refpa{eq:hol}, the holonomy transforms under gauge-transformations
by conjugation in $SU(N)$. Gauge-invariant
functions of the holonomy are therefore class functions $f$,
\[ f(h)=f(g hg^{-1}), \, \, \, \, \, \forall g\in SU(N). \]
A particular example of a class function is $T^0(n)$. Let us note some
properties of the conjugacy classes of $SU(N)$, the
classic source of information being \cite{hw:cg}. Any $SU(N)$ matrix is
conjugate to a diagonal matrix $D$. Two diagonal matrices
are conjugate if and only if their eigenvalues are related by permutation. Let
$D=\mbox{diag}(\lambda _1,\ldots ,\lambda _N)$.
Since $\det D=1$ we have,
\be
\lambda _N=\lambda _1^{-1} \cdots \lambda _{N-1}^{-1}.\label{eq:ln}
\ee
Furthermore, since $D$ is unitary the eigenvalues all have modulus $1$ i.e.
$\lambda _i=e ^{i \varphi _i}$, ($\varphi _i$ real
$i=1,\ldots ,N-1$). Any class function $f$ is therefore a function of $N-1$
eigenvalues, symmetric under permutations
\[ \lambda _i\leftrightarrow \lambda _j, \, \, \, \, \, i,j=1,\ldots ,N\]
where $\lambda _N$ is given by \refpa{eq:ln}, e.g. for $N=2$, $f(\lambda
_1)=f(\lambda _1^{-1})$. From now on, permutations
will always mean permutations of all $N$ eigenvalues, $\lambda _N$ being given
by \refpa{eq:ln}. We can express $T^0(n)$ in
terms of the eigenvalues of $h(x)$ (which are independent of $x$),
\be
T^0(n)=\lambda _1^n+\ldots +\lambda _{N-1}^n+\lambda _1^{-n} \cdots \lambda
_{N-1}^{-n}\label{t0l}.
\ee
When we've expressed $T^0(n)$ in terms of $\lambda$:s it is evident that all
the Mandelstam identities are satisfied.

\subsection{Momenta}
Having expressed $T^0(n)$ in terms of eigenvalues, it is natural to look for
some variables ``conjugate'' to the eigenvalues.
We do this by postulating the brackets,
\bea
\{ \lambda _i,p_j \} &=& i \delta _{ij} \lambda _i\label{lp}\\
\{ p_i,p_j \} &=& 0.
\eea
This defines the $N-1$ momenta $p_i$. We obviously already have the bracket,
\[ \{ \lambda _i,\lambda _j \}=0. \]
Furthermore, from \refpa{lp} follows $\{ \lambda _i^n,p_j \}=i n \delta _{ij}
\lambda _i^n $. Now assume that $T^1(n)$ is
purely first order in momenta (with $\lambda$-dependent coefficients). By
inserting this ansatz into \refpa{t1t0} using
\refpa{t0l} one finds (uniquely),
\be
T^1(n)=g \sum _{i=1}^{N-1}(\lambda _i^n-\frac{1}{N} T^0(n))p_i\label{t1pi}.
\ee
A check shows that this expression for $T^1(n)$ satisfies \refpa{t1t1}. Now
having found this, assume that $H$ is purely
second order in momenta. Inserting this ansatz into \refpa{htp} (for $p=0$) one
obtains the unique expression,
\be
H=\frac{g^2 L}{2}( \sum _{i=1}^{N-1}p_i^2-\frac{1}{N}(\sum
_{i=1}^{N-1}p_i)^2)\label{hpi}.
\ee
Having found \refpa{hpi} one can generate any $T^p(n)$ using \refpa{htp}. In
particular one finds,
\be
T^2(n)=g^2 \sum _{i=1}^{N-1}p_i^2 \lambda _i^n-\frac{g}{N}T^1(n) \sum
_{i=1}^{N-1}p_i-\frac{g^2}{N}\sum _{i=1}^{N-1} p_i \sum
_{i=1}^{N-1} p_i \lambda _i^n.
\ee
Checking, one finds that this expression for $T^2(n)$ satisfies \refpa{t2t0}
and \refpa{t2t1}. Let us finally see how
gauge-transformations affect the momenta. A gauge-transformation permutes the
eigenvalues. In particular exchanging
$\lambda _i$ and $\lambda _j$ exchanges $p_i$ and $p_j$ for \refpa{lp} to hold
after the transformation.
Changing $\lambda _i$ into $\lambda _N=\lambda _1^{-1} \cdots \lambda
_{N-1}^{-1}$ one finds $p_i\rightarrow -p_i$ and
$p_j\rightarrow p_j-p_i \, \, (j\neq i)$. As a consistency check one finds that
indeed $T^1(n)$ and $T^2(n)$ are
gauge-invariant.

\section{Quantization}
Poisson brackets without ordering problems will go over into commutators
unchanged,
\bea
\lbrack \hat T^0(n),\hat T^0(m) \rbrack &=& 0\label{eq:t0t0}\\
\lbrack \hat T^1(n),\hat T^0(m) \rbrack &=& \gh m (\hat
T^0(n+m)-\frac{1}{N}\hat T^0(n) \hat T^0(m))\label{eq:t1t0}\\
\
\lbrack \hat H,\hat T^p(n) \rbrack &=& \gh L n\, \hat
T^{p+1}(n)\label{eq:htp}\\
\lbrack \hat \lambda _i,\hat p _j \rbrack &=& -\hbar \delta _{ij} \hat \lambda
_i
\eea
Let these operators act on wavefunctions that are class functions, i.e.
symmetric functions of the $N-1$
eigenvalues of the holonomy $h(x)$. Hence let $\hat \lambda _i$ and $\hat p_i$
act as,
\bea
\hat \lambda _i \Psi (\lambda _1,\ldots ,\lambda _{N-1}) &=& \lambda _i \Psi
(\lambda _1,\ldots ,\lambda _{N-1})\\
\hat p_i \Psi (\lambda _1,\ldots ,\lambda _{N-1}) &=& \hbar \lambda _i \de
_{\lambda _i}
\Psi (\lambda _1,\ldots ,\lambda _{N-1}).
\eea
Under a $\Z _N$ transformation, $\Psi (\lambda _1,\ldots ,\lambda _{N-1})$
transforms into $\Psi (\xi \lambda _1,\ldots ,\xi
\lambda _{N-1})$, where $\xi ^N=1$.
Phasespace functions without ordering problems will turn into operators
unchanged i.e.,
\bea
\hat T^0(n) &=& \hat \lambda _1 ^n+\ldots +\hat \lambda _{N-1}^n+\hat \lambda
_1^{-n} \cdots \hat \lambda _{N-1}^{-n}\\
\hat H &=& \frac{g^2 L}{2}(\sum _{i=1}^{N-1}\hat p_i^2-\frac{1}{N}(\sum \hat
p_i)^2).
\eea
$\hat T^0(n)$ will obviously satisfy the Mandelstam identities now. Having
defined $\hat T^0$ and $\hat H$ we define
$\hat T^p$ for $p \geq 1$ by \refpa{eq:htp}. Hence one e.g. finds,
\be
\hat T^1(n)=g \sum _{i=1}^{N-1}(\hat \lambda _i^n-\frac{1}{N}\hat T^0(n))\hat
p_i+\gh n \frac{N-1}{2N}\hat T^0(n).
\ee
Comparing with the classical expression \refpa{t1pi} one sees that $\hat
T^1(n)$ has aquired a quantum correction.
Checking \refpa{eq:t1t0} it is found to be satisfied. One also finds,
\be
\lbrack \hat T^1 (n),\hat T^1(m) \rbrack =\gh (m-n) \hat T^1(n+m)+\frac{\gh
}{N}( n\hat T^0(m) \hat T^1(n)-m \hat T^0(n)
\hat T^1(m)).
\ee
Let us note some properties of $\hat H$. Introduce
\be
\Xi _{(n_1,\cdots ,n_{N-1})}(\{ \lambda \} )=\lambda _1^{n_1} \cdots \lambda
_{N-1}^{n_{N-1}},
\ee
where $n_1,\ldots ,n_{N-1}$ are integers and $\{ \lambda \}=(\lambda _1,\ldots
,\lambda _{N-1})$. $\Xi$ is an eigenvector of
$\hat H$, i.e.
\be
\hat H\, \Xi _{(n_1,\cdots ,n_{N-1})}(\{ \lambda \} )=\frac{ (\gh )^2 L}{2N}
P_N(n_1,\ldots ,n_{N-1})
\Xi _{(n_1,\cdots ,n_{N-1})}(\{ \lambda \} ),\label{eq:eigen}
\ee
where
\be
P_N(\{ n\} )=(N-1) \sum _{i=1}^{N-1} n_i^2-2\sum _{j>i=1}^{N-1} n_i n_j .
\ee

\subsection{Symmetric representation}
Let us investigate $\hat H$. This is in fact, in our formalism,
 the Hamiltonian derived in \cite{hh:ym}. The eigenstates are totally symmetric
linear combinations of $\Xi _{ ( \{ n \}) }$ (remember that physical states are
class functions), i.e.
\[ \Psi _{S(n_1,\ldots ,n_{N-1})} (\{ \lambda \} )=\sum _{perms} \Xi
_{(n_1,\ldots ,n_{N-1})}(\pi (\lambda _1),\ldots ,\pi (\lambda _{N-1})), \]
where $\pi$ permutes all $\lambda _i$:s including $\lambda _N$. Evidently, not
all indices $(n_1,\ldots ,n_{N-1})$ correspond to different
eigenstates. If we want these states to be $\Z _N$ invariant we have to require
$\sum _{i=1}^{N-1}n_i$ to be a multiple of
$N$. The eigenenergies are given by \refpa{eq:eigen}. The action of the loop
variables is very simple on the eigenstates, e.g.
\[
\hat T^0(n)\Psi _{S(n_1,\ldots ,n_{N-1})}(\{ \lambda \})=\sum _{i=1}^{N-1}\Psi
_{S(n_1,\ldots ,n_i+n,\ldots ,n_{N-1})}(\{ \lambda \})+
\Psi _{S(n_1-n,\ldots ,n_{N-1}-n)} (\{ \lambda \}).
\]
An inner product is determined by requiring $(\hat T^0 (n))^\da =\hat T^0(-n)$
and $\hat H^\da =\hat H$. Then \refpa{eq:htp}
implies that all the classical reality conditions, \refpa{eq:reality}, are
quantized exactly i.e. $(\hat T^p(n))^\da =\hat
T^p(-n)$. Hence, (up to an overall factor),
\be
< \Phi _S,\Psi _S>=\int d\varphi _1 \cdots d\varphi _{N-1} \Phi _S^* (\{
\varphi \} )\Psi _S(\{ \varphi \} ).
\ee
Here all integrals are taken from $-\pi$ to $\pi$ in the angles. Alternatively
we can integrate over the eigenvalues,
\[ \frac{d\lambda _i}{i \lambda _i}=d\varphi _i .\]
Different eigenstates are orthogonal using this inner product. The groundstate
is $\Psi _{S(0,\ldots ,0)}$ and it has zero energy.

\subsection{Antisymmetric representation}
Let's make a quantum canonical transformation using $C=\Delta ^{-1}$ where
$\Delta$ is,
\[ \Delta =\prod _{j>i=1}^N(\lambda _i-\lambda _j).\]
For a general discussion of such transformations see \cite{aa:ct}. An arbitrary
operator $\hat O$ will be mapped into
\[ \hat O'=C \hat O C^{-1}=\Delta ^{-1} \hat O \Delta .\]
We note that it is a well-defined canonical transformation mapping
gauge-invariant operators into gauge-invariant
operators. Under this transformation $\hat T^0$ is invariant while $\hat
H'=\Delta ^{-1} \hat H \Delta \neq \hat H$.
This is (up to a constant) the radial part of the Laplacian on $SU(N)$,
\cite{hel}, which is the Hamiltonian
considered in \cite{sgr:ym}. $\Delta$ is totally antisymmetric under
permutations of eigenvalues. Hence eigenstates of
$\hat H'$ are given as,
\[ \Psi _{A(n_1,\ldots ,n_{N-1})} (\{ \lambda \} )=\Delta ^{-1} \sum _{perms}
\mbox{sgn}(\pi)\,  \Xi _{(n_1,\ldots ,n_{N-1})}(\pi (\lambda _1),\ldots ,
\pi (\lambda _{N-1})). \]
These are the characters of $SU(N)$. Eigenenergies are still given by
\refpa{eq:eigen}. The groundstate is
$\Psi _{A(1,\ldots ,N-1)}$ with energy
\[ (\gh )^2L\frac{N}{24}(N^2-1).\]
The
spectrum of $\hat H'$ is a proper subset of that of $\hat H$. Hence these
Hamiltonians are clearly physically inequivalent. The action of
loop variables on eigenstates is the same as for the symmetric representation.
The inner product
is,
\[ < \Phi _A,\Psi _A>=\int d\varphi _1 \cdots d\varphi _{N-1} \Delta \Delta ^*
\Phi _A^* (\{ \varphi \} )\Psi _A(\{ \varphi \} ).\]
The measure density $\Delta \Delta ^*$ is the measure density induced by the
Haar-measure on the group. Note how utterly sensible it is
from the point of view
of the group, e.g. the conjugacy class
$\lambda _1=\ldots =\lambda _{N-1}=1$ consists of a single group element, the
unit matrix, in contrast to a generic conjugacy class having all
eigenvalues distinct which consists of a set of group elements forming a
submanifold of the group with non-zero dimension. Thinking about
the group it is natural to give a larger weight to this generic conjugacy class
than the unit element class. $\Delta$ does just this as it vanishes
on the unit element class. In general, the so called singular set which is the
set of conjugacy classes having not all eigenvalues distinct, has
Haar-measure zero ($\Delta$ is zero on this set).

\subsection{Generalities}
Having found that a canonical transformation can map us into an inequivalent
representation of the algebra of
gauge-invariant operators we might wonder if we can construct other
inequivalent transformations by the same means. Hence
let $\hat C(\{ \lambda \} )$ be some canonical transformation. Now an arbitrary
operator $\hat O$ will transform as,
\[ \hat O'=\hat C \hat O \hat C^{-1} .\]
But we require gauge-invariant operators to be mapped into gauge-invariant
operators. Hence $\hat C(\{ \lambda \})$ must
either be totally symmetric or totally antisymmetric, and eigenstates of the
transformed Hamiltonian will clearly
be equivalent to either the symmetric or the antisymmetric representation. We
can in principle allow for momentum
dependent canonical transformations as well, as long as we're careful with
their (possible) kernels, but this
will not lead to anything new. This quantization ambiguity is also discussed in
\cite{h:cq} from a completely
different point of view.

\subsection{Dirac quantization}
What we have done is more or less reduced phase-space quantization since we
consider (almost) only gauge-invariant
functions on the constraint surface when quantizing. It would be interesting to
compare this to Dirac quantization
where one solves the constraint on the quantum level. It turns out that in the
cases we have worked out
($N=2$ and $N$=3), this approach leads to results that are equivalent to the
reduced approach. Let us define
the smeared classical constraint,
\be
C_\omega =\frac{1}{g}\intll{0}{L}{x}\mbox{tr}(\omega (x) D_x E(x)),
\ee
where $D_x E(x)$ is given by \refpa{eq:gauss} and $\omega (x)=\omega ^a(x)
t^a$. Gauge-transformations are
generated by $C_\omega$ and,
\bea
\{ A(x),C_\omega \} &=& -\frac{1}{g}+i \lbrack \omega (x),A(x) \rbrack ,\\
\{ C_{\omega _1},C_{\omega _2} \} &=& i C_{ \lbrack \omega _1,\omega _2 \rbrack
}.
\eea
Upon quantization \refpa{poisson} turns into,
\be
\lbrack \hat A^a (x),\hat E^b (x) \rbrack =i\hbar \delta ^{ab}\delta
(x-y).\label{cc}
\ee
Now when quantizing the constraint $C_\omega$ we choose to order the $\hat
E^a$:s to the right (it is actually ordering
independent). Hence define $\hat C_\omega$ as,
\be
\hat C_\omega =\intll{0}{L}{x}\omega ^a(x)(\de _x\hat E^a(x)+i g\lbrack t^a,t^b
\rbrack \hat A^a(x) \hat E^b(x)).
\ee
Then one obtains,
\bea
\lbrack \hat A(x),\frac{1}{i\hbar} \hat C_\omega \rbrack &=& -\frac{1}{g}\de _x
\omega (x)+i \lbrack \omega (x),\hat A(x)
\rbrack ,\\
\lbrack \frac{1}{i\hbar}\hat C_{\omega _1},\frac{1}{i\hbar}\hat C_{\omega _2}
\rbrack &=& i \frac{1}{i\hbar}\hat C_{\lbrack
\omega _1,\omega _2 \rbrack },
\eea
as one should. Finite gauge-transformations are given by,
\be
\hat \Lambda _\omega =e^{-\frac{1}{i\hbar} \hat C_\omega}.
\ee
There are no ordering problems in the expressions for $T^0(n)$ and $H$. Hence,
\bea
\hat T^0(n) &=& \mbox{tr}(\hat h^n(x)),\\
\hat H &=& \frac{1}{2} \intll{0}{L}{x}\mbox{tr}(\hat E^2(x)).
\eea
Then we define the higher order loop variables when acting on physical states
by \refpa{eq:htp}. This
corresponds to choosing some particular ordering of the classical expressions.
Let us now choose the connection representation i.e.
states are wavefunctionals of connections $\Psi (A)$ and,
\bea
\hat A^a(x) \Psi (A) &=& A^a(x) \Psi (A),\\
\hat E^a (x) \Psi (A) &=& \frac{\hbar}{i}\frac{\delta \Psi (A)}{\delta A^a(x)}.
\eea
This is a representation of \refpa{cc}. Now act upon $\Psi (A)$ with a finite
gauge-transformation. One then finds,
\be
\hat \Lambda _\omega \Psi (A)=\Psi (A'),
\ee
where $A'(x)$ is given by \refpa{gtr} and $\Lambda (x)=\exp (i \omega (x))$.
The physical states should be
gauge-invariant i.e. $\Psi (A)=\Psi (A')$. But we already know of solutions to
this equation, the classfunctions
of the holonomy of $A$. Hence physical states are symmetric functions of $N-1$
eigenvalues of the holonomy $h(x)$ of
$A(x)$,
\[ \Psi _{phys}(A)=\Psi (\lambda _1,\ldots ,\lambda _{N-1}).\]
It is necessary but not sufficient (in this case) for the physical states to be
annihilated by the constraint i.e.
$\hat C_\omega \Psi (A)=0$. This is because two different such states
(satisfying $\hat C_\omega \Psi (A)=0$) might
be related by a finite gauge-transformation and the path in the space of
connections goes via states not being
annihilated by the constraint. If one then restricts ones attention only to
states being annihilated by the
constraint, this effect will never be seen. We obviously have,
\be
\hat T^0(n) \Psi (\{ \lambda \} )=T^0(n) \Psi (\{ \lambda \} ),
\ee
where $T^0(n)$ is given by \refpa{t0l}. Let us work out everything explicitly
for $N=2$. Then we have
$\Psi _{phys}(A)=\Psi (\lambda _1)$ where $\Psi (\lambda _1)=\Psi (\lambda
_1^{-1})$ and,
\be
\lambda _1^2-\lambda _1 \mbox{tr}(h(x))+1=0.
\ee
Differentiating this equation with respect to $A^a(x)$ and using
\refpa{eq:func} we obtain,
\be
\frac{\delta \lambda _1}{\delta A^a(x)}=i g \frac{\lambda _1}{\lambda
_1-\lambda _1^{-1}}\mbox{tr}(t^a h(x)).
\ee
Hence,
\be
\hat E^a(x) \Psi (\lambda _1)=\frac{\hbar}{i}\frac{\delta \Psi (\lambda
_1)}{\delta A^a(x)}=
\frac{\hbar}{i}\frac{\delta \lambda _1}{\delta A^a(x)}\de _{\lambda _1}\Psi
(\lambda _1)=
\gh \, \mbox{tr}(t^a h(x))\frac{\lambda _1}{\lambda _1-\lambda _1^{-1}}\de
_{\lambda _1}\Psi (\lambda _1).
\ee
We can know see explicitly that $\Psi (\lambda _1)$ is annihilated by the
constraint i.e.,
\[ \hat C_\omega \Psi (\lambda _1)=0.\]
Proceeding, we find,
\be
\hat H\Psi (\lambda _1)=\frac{(\gh )^2 L}{4}(\lambda _1^2 \de _{\lambda _1}^2
+\frac{3\lambda _1+
\lambda _1^{-1}}{\lambda _1-\lambda _1^{-1}}\de _{\lambda _1})\Psi (\lambda
_1),\label{hsu2}
\ee
having used,
\[ \frac{\delta \mbox{tr}(t^a h(x))}{\delta A^a(x)}=ig\frac{3}{2}(\lambda
_1+\lambda _1^{-1}),\]
and
\[ \mbox{tr}(t^a h(x))\, \mbox{tr}(t^a h(x))=\frac{1}{2}(\lambda _1-\lambda
_1^{-1})^2.\]
The expression \refpa{hsu2} doesn't look very transparent but in fact,
\be
\hat H\Psi (\lambda _1)=\frac{(\gh )^2}{4}(\Delta ^{-1} (\lambda _1 \de
_{\lambda _1})^2 \Delta -1)\Psi
(\lambda _1),
\ee
where $\Delta =\lambda _1-\lambda _1^{-1}$ i.e. $\hat H$ is identical to the
Hamiltonian in the antisymmetric
representation up to a constant (unobservable) term. Add this term for complete
equivalence i.e. let
\[ \hat H=\frac{1}{2}\intll{0}{L}{x}\mbox{tr}(\hat E^2(x))+\frac{(\gh
)^2}{4}.\]
This ensures that $T^2(n)$, being defined by \refpa{eq:htp}, is continuous in
$n=0$.
Hence if we do a canonical transformation $C(\lambda _1)=\Delta$ we can map all
operators into the symmetric
representation. In particular $\hat E^a(x)$ transforms as,
\[ \hat E'^a (x)=C(\lambda _1) \hat E^a(x) C(\lambda _1)^{-1}=\hat E^a(x)-\gh
\frac{\lambda _1+\lambda _1^{-1}}
{(\lambda _1-\lambda _1^{-1})^2} \mbox{tr}(t^a h(x)). \]
This seems to be a very contrived representation for the electric field
operator and we cannot help getting a
feeling that somehow the antisymmetric representation is the ``correct''
representation. We have also checked
$N=3$. Then we again find that the Hamiltonian is given by the Hamiltonian in
the antisymmetric representation
except for the ground state energy but the calculations involved are much
longer than for $N=2$. We are
completely convinced that the same thing will happen for a general $N$, but we
haven't proved this.

\subsection{Loop representations}
Let us now try to establish a link between our representation of the
$T$-algebra and a loop representation. It has been
suggested \cite{rs:qgr} that one can go from the connection representation
(wavefunctionals of connections) to the
loop representation using the loop transform. Denoting a state in the
connection representation by $\Psi (A)$ it
looks like (in any dimension),
\[ \tilde \Psi (\gamma )=\int {\cal {D}}A\, \, T^0(\gamma ,A) \Psi (A),\]
where $\gamma$ is a loop and $T^0(\gamma ,A)=\mbox{tr} {\cal {P}}\exp (\oint
_\gamma A)$. In our case this would
translate into something like,
\[ \tilde \Psi (n)=\int \prod _{i=1}^{N-1} d\lambda _i \mu (\{ \lambda \} )
T^0(n) \Psi (\{ \lambda\} ),\]
where $\mu (\{ \lambda\} )$ is some kind of measure. This is clearly inadequate
(at least when $N>2$). $\tilde \Psi (n)$
 takes values on a single integer $n$ while $\Psi (\{ \lambda\} )$ takes values
on $N-1$ eigenvalues. Let instead
$\Psi _{(n_1, \ldots, n_{N-1})}(\{ \lambda\} )$ be the (complete) set of
eigenstates of $\hat H$ (i.e. $\hat T^2(0)$),
which are either equivalent to the eigenstates $\Psi _S$ in the symmetric
representation or the eigenstates $\Psi _A$ in
the antisymmetric representation. Then define the loop representation by means
of the transform,
\be
\tilde \Phi (n_1,\ldots ,n_{N-1})=< \Psi _{(n_1, \ldots, n_{N-1})}(\{ \lambda\}
),\Phi (\{ \lambda \} )>,
\ee
i.e.
\[ \tilde \Phi (n_1,\ldots ,n_{N-1})=\inner{n_1,\ldots ,n_{N-1}}{\Phi},\]
where $\inner{\{ \lambda \}}{n_1,\ldots ,n_{N-1}}=\Psi _{(n_1, \ldots,
n_{N-1})}(\{ \lambda\} )$. Furthermore define,
\be
\hat O\tilde \Phi (n_1,\ldots ,n_{N-1})=<\Psi _{(n_1, \ldots, n_{N-1})}(\{
\lambda\} ),\hat O\Phi (\{ \lambda \} )>=
<\hat O^\da\Psi _{(n_1, \ldots, n_{N-1})}(\{ \lambda\} ),\Phi (\{ \lambda \}
)>,
\ee
where $\hat O$ is any (gauge-invariant) operator. Hence we get,
\bea
\hat T^0(n) \tilde \Phi (n_1,\ldots ,n_{N-1}) &=& <\hat T^0(-n) \Psi _{(n_1,
\ldots, n_{N-1})}(\{ \lambda\} ),\Phi (\{ \lambda \} )>\nn \\
&=& \sum _{i=1}^{N-1}\tilde \Phi (n_1,\ldots ,n_i-n,\ldots ,n_{N-1})+\nn \\
&& \tilde \Phi (n_1+n,\ldots, n_{N-1}+n),
\eea
since
\[ \hat T^0(n) \Psi _{(n_1, \ldots, n_{N-1})}(\{ \lambda\} )=\sum
_{i=1}^{N-1}\Psi _{(n_1, \ldots, n_i+n,\ldots ,
 n_{N-1})}(\{ \lambda\} )+\Psi _{(n_1-n, \ldots, n_{N-1}-n)}(\{ \lambda\} ),\]
independent of representation. In particular, for $N=2$,
\[ \hat T^0(n)\tilde \Phi (n_1)=\tilde \Phi (n_1+n)+\tilde \Phi (n_1-n),\]
and we have the even and odd loop representation, arising from the symmetric
and antisymmetric representation respectively,
\[ \tilde \Phi _E(-n_1)=\tilde \Phi _E(n_1) \, \, \, \, \, \, \tilde \Phi
_O(-n_1)=-\tilde \Phi _O(n_1).\]
The action of $\hat T^0(n)$ is exactly what one expects and we see that the
even loop representation corresponds to the
ordinary loop representation while the odd one is something new. Let us also
see how $T^1(n)$ acts in the loop
representation. Using \refpa{eq:htp} for $p=0$,
\[ T^1(n)=\frac{1}{\gh Ln}\lbrack \hat H,\hat T^0(n)\rbrack ,\, \, \, \, \, (n
\neq 0),\]
and $T^1(0)=0$. Hence, using \refpa{eq:eigen},
\bea
\lefteqn{\hat T^1(n)\tilde \Phi (n_1,\ldots ,n_{N-1})=}\nn \\
&&-\frac{1}{\gh Ln}<\lbrack \hat T^0(-n),\hat H\rbrack \Psi _{(n_1,
\ldots, n_{N-1})}(\{ \lambda\} ),\Phi (\{ \lambda \} )>=\nn \\
&& -\frac{\gh }{N}( \sum _{i=1}^{N-1}(Nn_i-s)\tilde \Phi (n_1,\ldots
,n_i-n,\ldots ,n_{N-1})-s \tilde \Phi (
n_1+n,\ldots ,n_{N-1}+n))+\nn \\
&& \frac{\gh n(N-1)}{2N}\hat T^0(n)\tilde \Phi (n_1,\ldots ,n_{N-1}),
\eea
where $s=\sum _{i=1}^{N-1}n_i$. In the special case $N=2$ this expression
becomes,
\be
\hat T^1(n)\tilde \Phi (n_1)=-\frac{\gh}{2}(n_1\tilde \Phi (n_1-n)-n_1 \tilde
\Phi (n_1+n))+\frac{\gh n}{4}\hat T^0(n)
\tilde \Phi (n_1).
\ee
This is what to be expected from the ordinary loop representation except for
the last term which contains a $\hat T^0$ and
an overall sign.
This term has arisen since we are doing ``minimal quantization'' i.e. if there
aren't any ordering problems in a
classical Poisson bracket we let this bracket go into a commutator without any
modifications. This is not the
philosophy adopted in \cite{rs:qgr}, giving the reason for the discrepancy. The
sign is there because contrary to the ordinary
loop representation, operators act directly on loop states instead of first
acting on a state and then evaluating that state
on a particular loop. It should be noted that it is
important for applications to general relativity exactly what one chooses as
$\hat T^1$ and $\hat T^2$ since the
diffeomorphism and hamiltonian constraint respectively are defined as limits of
those operators.

\section{Conclusion}
We investigate the Poisson bracket algebra of $T$-variables and show that on
the constraint surface the $T$-variables can be expressed
as functions of the eigenvalues of the holonomy and their associated momenta.
The essential structure of the $T$-algebra is seen
to be very simple, with $T^2(0)$ acting as a kind of generating function.
Quantization then proceeds without problems, the
only surprise being that a canonical transformation can map us into an
inequivalent representation of the quantum $T$-algebra.
We then compare this reduced phase space approach to Dirac quantization and
find it to give essentially equivalent results.
Dirac quantization seems however to ``naturally'' prefer one of the
inequivalent representations. Having the complete set
of eigenstates of the Yang-Mills Hamiltonian (essentially $\hat T^2(0)$) one
can define a loop transform
and hence loop representations. The main conclusion is then that the loop
representation isn't essential for the quantum $T$-algebra.
Rather, by using a more solid starting point, the loop representation might or
might not be derived from that formalism depending on
whether it is ``complete'' or not. There is really just one point which isn't
clear, should the symmetric or the antisymmetric
representation of the quantum $T$-algebra be preferred somehow? As we have
seen, the antisymmetric representation seems much
more natural from the point of view of Dirac quantization and as long as we're
only considering pure gauge theory we can say
no more. Coupling fermions to the theory might give further insight into this
problem.
\\
We wish to thank B.E.W. Nilsson and B.S. Skagerstam for discussions.

\end{document}